\documentstyle[12pt,cite,fleqn]{article}
\def\@cite#1#2{\textsuperscript{[{#1\if@tempswa , #2\fi}]}}
\makeatother \textheight 25.0cm \textwidth 17cm \topmargin -2.5cm
\input epsf
\title{\Large{\bf{Multi-deformed Configurations and Shape Coexistence for Superheavy
Elements  }}\thanks{Supported by National Natural Science
Foundation of China (10275037) and Specialized Research Fund for
the Doctoral Program of Higher Education of China (20010055012)}}

\author{ \footnotesize X.H. Zhong,  \footnote {E-mail: zhongxianhui@mail.nankai.edu.cn}
\, \,L. Li,  \footnote {E-mail: lilei@nankai.edu.cn}  \,\, P.Z.
Ning \footnote {E-mail: ningpz@nankai.edu.cn}
\\\footnotesize  \emph{Department of Physics, Nankai University,
 Tianjin 300071, P. R. China}}
\begin{document}
\maketitle

\begin{abstract}

We use the considered axial deformed relativistic mean field
theory to perform systematical calculations for $Z=112$ and 104
isotopic chains with force parameters NL3, NL-SH and NL-Z2 sets.
Three deformed chains (oblate, moderate prolate and super-deformed
chain) are found for $Z=112$ and 104 isotopic chains. It is found
that there is a chain of super-deformed nuclei which can increase
the stability of superheavy nuclei in the $Z=112$ isotopic chain.
Shape coexistence is found for $Z=112, 104$ isotopic chain and the
position is defined. For moderate prolate deformed chains of
$Z=112$ and 104, there is shell closure at $N=184$ for moderate
prolate deformed chain. For oblate deformed chain of $Z=112$, the
shell closure appears around at $N=176$. For super-deformed chains
of $Z=112$ and 104, the position of shell closure have strong
parameter dependence. There is shell anomalism for oblate or
superdeformed nuclei.


\end{abstract}
\section{Introduction}
   Since the possible existence of superheavy elements was predicted
   in 1960s by nuclear theoreticians, the search of superheavy
   elements in nature has become a hot topic for
   scientists. Empirically, many heavy elements were identified by nuclear synthesis.
   At first, the elements $Z=105\sim108$ were successfully produced and both physicists and chemists
   agree on the existence of these elements although their
   half-lives are not very long.  During $1995\sim1996$, because of
   the participating of more and more large laboratories in researches of new elements,
   the elements $Z=110\sim112$ were
   produced by Hofmann et al. at GSI in Germany\cite{1,2,3}. $Z=114$ was produced by Oganessian et
   al. at Dubna in Russia in 1999\cite{4,5}. One year later it was again reported that
   $Z=116$ was synthesized at Dubna\cite{pp}. Recently,
   $^{288}115$ and $^{287}115$ were synthesized at FLNR,
   JINR\cite{pp1}.
   The achievement in producing new elements speeds up the
   researches on superheavy nuclei not only in experiment  but also
   in theory. Theoretically, several models have been  used
   and many aspects such as the collisions,
   structure, and stability  have been
   investigated \cite{p1,p2,p3,p4,p5,30,31,32,33,34,35} for heavy and superheavy
   elements.

   Recently, there are systematically calculations for superheavy elements within the framework
   of deformed relativistic mean-field (RMF) theory \cite{d1,d2,d3}.
   These calculations predicted that there are shape
   coexistence and super-deformation in the ground state of superheavy
   nuclei and deformation can be an important cause for the stability
   of superheavy nuclei based on a constraint RMF calculation. The
   conclusion changes the usual conception  that the existence
   of superheavy elements is due to their spherical shell
   structure and makes people recognize  that deformed configurations are as
   important as the spherical one for stability of superheavy
   elements in theory.

   However, the constraint RMF calculation
   consumes much time, thus the calculation is very limited.
   Although there are many systematical
   calculations for superheavy nuclei which find there are a
   moderate prolate solution, an oblate solution, and a superdeformed
   prolate solution for the same element\cite{d2,d3}, there is no
   calculation which can give out deformed configurations of a whole isotopic
   chain.  No information tells us that whether there are two or more
   deformed configurations for most of the  nuclei in a whole isotopic
   chain or only exist in a few nuclei in the isotopic chain. Our aim is to
   study these properties in a long isotopic chain.
   Thus, we select $Z=104$ and $Z=112$ isotopic chains and carry out
   systematical calculation within the framework of deformed
   relativistic mean-field (RMF) theory. The binding energy,
   quadrupole deformation, root mean square radii, shape
   coexistence and shell closure are the investigative projects.

\section{The formalism of the relativistic mean-field theory}
The relativistic mean-field theory has been widely used to
describe finite nuclei, in the RMF method, the local Lagrangian
density is given as\cite{c1,c2}
\begin{eqnarray}
 {\mathcal{L}}=   \bar{\psi} (
i\gamma^{\mu}\partial_{\mu}- M)\psi- g_{\sigma} \bar{\psi}
\sigma\psi-
 g_{\omega}\bar{\psi}\gamma^{\mu}\omega_{\mu}\psi- g_{\rho}
 \bar{\psi }\gamma^{\mu}\rho^{a}_{\mu}\tau^{a}\psi\cr+
\frac{1}{2}\partial^{\mu}\sigma\partial_{\mu}\sigma-\frac{1}{2}m
_{\sigma}^{2}\sigma^{2}- \frac{1}{3}g
_{2}^{2}\sigma^{3}-\frac{1}{4} g _{3}^{2}\sigma^{4}\cr-
\frac{1}{4}\Omega^{\mu\nu}\Omega_{\mu\nu} +\frac{1}{2}
m_{\omega}^{2}\omega^{\mu}\omega_{\mu} -\frac{1}{4}  R ^{a \mu
\nu}   R _{\mu\nu}^{a}+
 \frac{1}{2} m_{\rho}^{2}   {\rho^{a\mu}}  {\rho
 ^{a}_{\mu}}
\cr-\frac{1}{4} F^{\mu\nu}F_{\mu\nu}-e \bar{\psi }\gamma^{\mu}
A^{\mu}\frac{1}{2}(1-\tau^{3})\psi.
\end{eqnarray}
The meson fields included are the isoscalar  $\sigma$ meson, the
isoscalar-vector $\omega$ meson and the isovector-vector $\rho$
meson. $M$,  $m_{\sigma}$, $m_{ \omega}$ and $m_{\rho}$ are the
nucleon-, the $\sigma$-, the $\omega$- and the $\rho$-meson
masses, respectively, while $g_{\sigma}$, $g_{\omega}$, $g_{\rho}$
and $e^{2}/4\pi=1/137$ are the corresponding coupling constants
for the mesons and the photon. The isospin Pauli matrices  are
written as $\tau^{a}, \tau^{3}$ being the third component of
$\tau^{a}$. The field tensors of the vector mesons and of the
electromagnetic fields take the following form:
\begin{eqnarray}
 \Omega^{\mu\nu}&=&\partial^{\mu}\omega^{\nu}-\partial^{\nu}\omega^{\mu},\cr
 R ^{a\mu\nu}&=&\partial^{\mu}\rho^{a\nu}-\partial^{\nu}\rho^{a\mu},\cr
 F ^{\mu\nu}&=&\partial^{\mu}A^{\nu}-\partial^{\nu}A^{\mu}.
\end{eqnarray}

The variational principle gives the equations of motion. For the
static case the meson fields and photon field operators are
assumed to be classical fields and they are time independent. They
are replaced by their expectation values. The symmetries of the
system simplify the calculations considerably. In all the systems
considered in this work, there exists time reversal symmetry, so
there are no currents in the nucleus and therefore the spatial
vector components of $\omega^{\mu}$, $\rho^{a\mu}$  and $A^{\mu}$
vanish. This leaves only the time-like components, $\omega^{0}$,
$\rho^{a0}$ and  $A^{0}$. Charge conservation guarantees that only
the 3-component of the isovector $\rho^{00}$ survives. Finally we
have the following Dirac equation for the nucleon:

\begin{equation}
  \{-i \alpha \nabla+V(r)+\beta[M+S(r)]\}\psi_{i}
  =\varepsilon_{i}\psi_{i},
\end{equation}
where $V(r)$ is the vector potential
\begin{equation}
V(r)=g_{\omega}\omega^{0}(r)+g_{\rho}\tau^{3}\rho^{00}+e\frac{1+\tau^{3}}{2}A^{0}(r),
\end{equation}
and $S(r)$ is the scalar potential
\begin{equation}
S(r)=g_{\sigma}\sigma(r).
\end{equation}
The Klein-Gordon equations for the  mesons and the electromagnetic
fields with the densities as sources are
\begin{eqnarray}
   \{-\Delta+m_{\sigma}^{2}\} \sigma(r)&=&-\:g_{\sigma}\rho_{s}(r)-g_{2}\sigma^{2}(r)-g_{3}\sigma^{3}(r),\\
   \{-\Delta+m_{\omega}^{2}\} \omega_{0}(r)&=&g_{\omega}\rho_{v}(r),\\
   \{-\Delta+m_{\rho}^{2}\} \rho_{00}(r)&=&g_{\rho}\rho_{3}(r),\\
   -\Delta A^{0}(r)&=&e\rho_{c}(r).
 \end{eqnarray}
The corresponding densities are
\begin{eqnarray}
\rho_{s}(r)&=&\sum_{i=1} ^{A}\overline{\psi}_{i}(r)\psi_{i}(r) ,\\
\rho_{v}(r)&=&\sum_{i=1} ^{A} \psi^{\dagger} _{i}(r)\psi_{i}(r) ,\\
\rho_{3}(r)&=&\sum_{i=1} ^{A} \psi^{\dagger} _{i}(r)\tau^{3}\psi_{i}(r) ,\\
\rho_{c}(r)&=&\sum_{i=1} ^{A}
\psi^{\dagger}_{i}(r)((1-\tau^{3})/2)\psi_{i}(r) .
\end{eqnarray}

Now we have a set of coupled equations for mesons and nucleons and
they will be solved consistently by iterations.

\section{Numerical calculation and analysis}
The validity of deformed relativistic mean-field (RMF) theory in
the calculation for superheavy nuclei is tested in previous
papers\cite{d1,d2,d3}, so we do not test the validity any more in
our work. In the process calculation, three typical sets of  force
parameters NL3\cite{q1}, NL-SH\cite{q2}, and NL-Z2\cite{p3} in RMF
model are chosen. The method of harmonic basis expansions is used
in solving the coupled RMF equations. The number of bases is
chosen as $N_{f}=12, N_{b}=20$. Pairing has been included using
the BCS formalism. In the BCS calculations we have used constant
pairing gaps $\Delta_{n}=\Delta_{p}=11.2/\sqrt{A}$ MeV\cite{f1}.
This input of pairing gaps is used in nuclear physics for many
years. Although the BCS model may fail for light neutron-rich
nuclei, the nuclei studied here are not light neutron-rich nuclei
and the RMF results with BCS treatment should be reliable. The
different inputs of $\beta_{0}$ lead to different iteration
numbers of the self-consistent calculation and different
computational time, but physical quantities such as the binding
energy and the deformation do not change much, which is tested in
Ref.\cite{d2}. Thus, when we carry out calculation, we only choose
a proper initial $\beta_{0}$ and neglect its effect on our
results.

\subsection{\emph{Quadrupole deformation} }
The quadrupole deformation of isotopic chain $Z=112$, $160\leq N
\leq 200$ with the force parameters NL3, NL-SH and NL-Z2 are
listed in figure 1.  There are three deformed chains 1, 2 and 3,
denoted with circle, triangle and star respectively, which can be
seen for all the force parameters NL3, NL-SH and NL-Z2 in figure
1.  1 is a oblate chain, the quantities of quadrupole deformation
$\beta_{2}\leq-0.3$.  2 is a moderate or light prolate deformed
chain except that there is light oblate deformation in the NL-Z2
calculation when $N\geq186$. It is very interesting that
 3 is  a super-deformed chain with $\beta_{2}\geq 0.4$ except
$\beta_{2}$=0.34, 0.33, 0.32 for $N$=196, 198, 200 in the NL-SH
calculation, but they  are still around $\beta_{2}$=0.4. The
phenomenon of several deformed configurations for a single
superheavy element is predicted in Refs.\cite{d2,d3} with
constraint RMF calculation. In our work, it is the first time to
obtain several deformed chains in a isotopic chain without using
any constraint RMF calculations at all.

Is it a general phenomenon that there are several deformed chains
for a isotopic chain of superheavy elements? To answer this
question, we also perform RMF calculation for $Z=104$, $152\leq N
\leq 198 $ isotopic chain. The results are shown in figure 1 as
well. All the results of the force parameters (NL3, NL-SH, and
NL-Z2) show three deformed chains (oblate, moderate prolate and
super-deformed prolate deformation ) for the $Z=104$ isotopic
chain as well as $Z=112$. Although there are three deformed bands
for the $Z=104$ isotopic chain, the super-deformed configurations
appear with neutron number $N\geq 168$ (for NL3 and NL-Z2
calculation) or $N\geq 170$ (for NL-SH calculation). The oblate
deformation appears in the region of $N \leq 168$ for NL3
calculation, $N \leq 170$ for NL-SH calculation, however, it
appears in the whole region of $Z=104$ isotopic chain for the
NL-Z2 calculation. Why  oblate deformation does not appear in the
whole region for NL3 and NL-SH calculations?  Oblate deformed
configurations which are predicted by NL-Z2 set but not by NL3 and
NL-SH sets in the region. Are they physical solutions? In figure
1, we see there is a sudden change at $N=176$ in the oblate chain
for the NL-Z2 calculation, when $N\geq 178$ the oblate deformation
changes gradually with the variation of the neutron number. It
indicates that the results of NL-Z2 calculation in the region
$N\geq 178$ are physical solutions. The NL-Z2 set may be better
than the other parameters in  describing the deformation in larger
neutron region. From the above mentioned analysis, we conclude
that it is a general phenomenon that there are several deformed
configurations for a isotopic chain of superheavy elements, and
that the super-deformed nuclei can exist in the superheavy
elements. Since there are several deformed configurations for the
superheavy elements, is there shape coexistence in a superheavy
element? This phenomenon will be discussed in the subsection 3.3
in detail.

From figure 1, we can see there are minima or maxima in the
deformed chains. In super-deformed chain of $Z=112$, the minimum
appears at $N=172$ for NL3, $N=168$ for NL-SH and $N=176$ for
NL-Z2, and the maximum  appears at $N=182$ for all the force
parameters. And in super-deformed chain of $Z=104$, there is no
minima for NL3; for NL-SH, the minimum is at $N=188$; for NL-Z2,
sudden changes occur at $N=182,184$, which is the very minimum. In
the moderate deformed chain of both $Z=112$ and $Z=104$, the
minimum appears at around $N=184$ for all the three parameters. In
oblate chain of $Z=112$, there is a maximum at $N=176$ for all the
force parameters. There is no obvious kink for NL3 and NL-SH
calculation in oblate chain of $Z=104$, but there is a large peak
at $N=176$ and a small peak at $N=190$ for NL-Z2 calculation.
Generally the minimum of the prolate deformed chain and the
maximum or peak of oblate deformed chain correspond to shell
closure. In the moderate deformed chain, the minimum
($\beta_{2}\approx 0$) appears at $N\simeq184$, which agrees with
the prediction in refs.\cite{g1,g2,g3,g4,g5,p3} that spherical
neutron shell closures occur at $N=184$. For superdeformed heavy
nuclei the shell closures depend on the force parameters. There is
strong trend that shell closure occurs at $N=176$ for $Z=112$ in
the oblate deformed chain, for all the parameters giving a peak at
$N=176$. If the minimum or maximum is the sign of shell closure,
the above analysis indicates that the positions of shell closure
for the super-deformed and oblate deformed nuclei are different
from that of spherical nuclei and have strong parameter
dependence.  We believe that there may be shell anomalism in
oblate and super-deformed prolate superheavy  nuclei. In fact, the
phenomenon of shell anomalism is predicted in some
Refs\cite{k1,k2,k3}.

\subsection{\emph{Root mean square radii (rms)}}
The root mean square (rms) radii are the project of investigation,
for they contain  a lot of important information of ground state
properties. In figure 2,  the rms radii of oblate, moderate
prolate and super-deformed prolate configurations with NL3, NL-SH
and NL-Z2 for the isotopic chains of $Z=112, 104$ are listed. The
solid symbol stands for neutron and empty symbol stands for proton
radii. The circles, up triangles and down triangles denote the
oblate, moderate prolate and super- deformed configuration
respectively.

In figure 2, it is seen that there is a gap $0.15\sim 0.38 fm$ for
$Z=112$ and $0.17\sim 0.45 fm$  for $Z=104$ isotopic chain between
neutron rms radii and proton rms radii of the same deformed
configuration for all the force parameters. The gap become larger
and larger with the increasing of the neutron number. For the
three parameters NL3, NL-SH and NL-Z2, if we select the rms radii
of NL3 as the standard, the NL-SH calculation underestimates the
rms radii about $0.05 fm $ and  the NL-Z2 calculation
overestimates the rms radii about $0.1 fm$  in  the same deformed
configuration for both $Z=112$ and $Z=104$ isotopes. We also can
see three obvious chains of rms radii  for the three deformed
configurations respectively in figure 2.

For $Z=112$ ($160\leq N\leq 200$) isotopic chain, NL3, NL-SH and
NL-Z2 calculations show that the rms radii of moderate prolate
deformed configuration are the smallest among  the three deformed
ones, and the rms radii of super-deformed one are the largest in
the region $N \leq 190$, namely, nuclei with large absolute values
$\beta_{2}$ tend to have larger rms radii in this region, however,
when $N \geq 192$ the rms radii of oblate deformation are the
largest. The gap of rms radii between the different deformed
configurations is about $0.1\sim 0.2 fm$ in general. For moderate
prolate deformed configuration, there are obvious kinks in the
proton rms radii at $N=184$, which agrees with the results  in
subsection 3.1 that there is  shell closure at N=184.  For oblate
deformed one, the calculation of NL3, NL-SH and NL-Z2 shows kink
in the proton rms radii at $N=176$, but the NL-Z2 calculation is
not very obvious. Together with the prediction of subsection 3.1,
we are convinced that there is shell closure at $N=176$ for oblate
deformed configuration. Finally, for super-deformed configuration,
the NL3 and NL-SH calculations show kinks  at $N \simeq 168, 184$
in the proton rms radii, and the kink appears at $N \simeq 190$
for NL-Z2, there also are obvious kinks at $N\simeq 186$ for NL3
and NL-SH at $N\simeq 192$ in proton rms radii. In summary, the
characters of shell closure for super-deformed configuration are
not very obvious in the region $N<184$, and have strong parameter
dependence. There are some sudden changes at $N=198,200$ for
moderate prolate deformed configuration, at $N=196, 198, 200$ for
super-deformed one with NL-SH calculation and  at $N=198,200$ for
oblate deformed configuration with NL-Z2 calculation. The
anomalism maybe come from the validity of the force parameter in
these regions.

For $Z=104$ ($152\leq N\leq 198$) isotopic chain, figure 2 shows
that the rms radii of  moderate prolate deformed configuration are
larger than those of oblate one for both NL3 and NL-SH
calculations. The gap between them is about $0.03\sim 0.07 fm$.
For NL-Z2 calculation, it scarcely shows any  gap in the rms radii
between moderate prolate and oblate deformed configuration when
$152\leq N \leq 176$ in figure 2. All the calculations of NL3,
NL-SH and NL-Z2 indicate that the radii of super-deformed
configuration are larger than that of the moderate. The radii of
oblate deformed configuration for NL-Z2 in the region $N\geq 178$
lie  between the radii of super-deformed configuration and the
moderate one. When $N\geq 186$, it is very close to the
super-deformed configuration's. As a whole, nuclei with large
absolute values $\beta_{2}$ tend to have larger rms radii.  In
figure 2,  obvious kinks can be seen at $N=184$ in the proton rms
radii for moderate deformed configuration of all the parameters,
which consists with the prediction in subsection 3.1 that $N=184$
is a magic number and shell closure exists there. Then we see the
results for superdeformed configuration in figure 2. The rms radii
suddenly become small at $N=172$ with NL3 calculation. $N=172$ is
a magic number for spherical nuclei in many Refs.\cite{p1,p3}. Is
$N=172$ still a magic number for superdeformed configuration, and
does the sudden change come from the shell closure at $N=172$ as
well? It is very difficult to answer the question, because the
other two force parameters NL-SH and NL-Z2 can not reproduce that
phenomenon. There is also sudden change at $N=194$ with NL3
calculation. The NL-SH calculation shows that the rms radii change
gradually with the variation of neutron number and there is a kink
at $N=176$, from which the radii increases much more slowly with
the increasing of neutron. It is seen  that the rms radii suddenly
become small at $N=182$ and 184 with NL-Z2 calculation. Is it
caused by the shell closure at around $N=184$? However, it is not
predicted by NL3 and NL-SH calculations. Finally, let's see the
results of oblate deformed configuration. We only list the results
in the region $152\leq N \leq 168$ for NL3 and $152\leq N \leq
170$ for NL-SH, because when we perform calculation with NL3 at
$N=170$ and with NL-SH at $N=172$, the results suddenly change to
moderate deformed configuration's. The calculation with NL-Z2 set
shows there is sudden change at N=176. In summary, it is a very
strange region in $N=168\sim 178$, there maybe exist complicate
shell structure.

From the analysis for the rms radii of  $Z=112$ and 104 isotopes,
we find that there is strong parameter dependence in predicting
the position of shell closure for oblate and super-deformed
configuration. Shell anomalism maybe occur in the oblate and
super-deformed configuration.

\begin{table}[h]
\begin{center}
 \caption{\footnotesize Binding energies  for $Z=112$  isotopic
chains with calculation of NL3, NL-SH and NL-Z2 sets. $B_{1},
B_{2}$ and $B_{3}$ denote the binding energies of oblate, moderate
prolate and super-deformed prolate configurations. $N$ is the
neutron number. } \label{} \vspace{0.1cm}\scriptsize
\begin{tabular}{|c|c|c|c|c|c|c|c |c|c|c|}\hline\hline
 &\multicolumn{3}{c|}{NL3}&\multicolumn{3}{c|}{NL-SH}&\multicolumn{3}{c|}{NL-Z2}\\
\hline
 {$N$ } &  $B_{1}$(MeV)& $B_{2}$ (MeV)& $B_{3}$(MeV) & $B_{1}$(MeV) & $B_{2}$(MeV) &$B_{3}$(MeV)&$B_{1}$(MeV)& $B_{2}$(MeV)&$B_{3}$(MeV)\\
\hline
160&1946.55&1958.84&1953.41&1948.74&1962.90&1955.63&1946.00&1955.60&1952.48\\
162&1961.41&1973.53&1968.65&1963.73&1977.47&1970.63&1960.83&1970.81&1967.61\\
164&1975.69&1987.53&1982.56&1978.47&1991.75&1984.57&1975.06&1984.51&1981.87 \\
166&1989.47&2001.13&1995.92&1992.56&2005.42&1998.00&1988.91&1997.96&1995.55\\
168&2002.88&2014.09&2009.13&2005.83&2017.59&2011.23&2002.49&2011.19&2008.79\\
170&2015.93&2026.18&2022.19&2018.61&2028.06&2024.43&2015.78&2023.79&2021.84\\
172&2028.60&2037.76&2035.05&2031.03&2040.07&2037.43&2028.69&2036.09&2034.51\\
174&2040.84&2049.03&2047.34&2043.28&2051.35&2049.84&2041.13&2047.98&2046.75\\
176&2052.75&2059.90&2058.66&2054.83&2062.30&2060.72&2053.22&2058.94&2058.62 \\
178&2064.00&2070.54&2069.42&2065.97&2072.74&2070.93&2065.05&2070.14&2070.03\\
180&2074.84&2080.39&2079.65&2076.35&2082.60&2080.87&2076.62&2080.34&2080.93\\
182&2085.21&2090.59&2089.50&2086.11&2092.27&2090.59&2087.72&2090.29&2091.18\\
184&2094.79&2100.68&2099.03&2095.33&2101.54&2099.97&2097.96&2100.84&2100.90 \\
186&2103.90&2108.92&2108.07&2104.13&2109.80&2108.89&2107.68&2109.50&2110.45\\
188&2112.76&2116.72&2116.79&2112.56&2118.12&2117.82&2117.09&2117.75&2119.70\\
190&2121.26&2124.88&2125.48&2120.50&2126.36&2126.67&2126.22&2125.67&2128.76\\
192&2129.42&2132.67&2134.31&2128.10&2134.23&2135.18&2135.07&2133.36&2137.73 \\
194&2137.31&2139.97&2142.95&2135.52&2141.62&2143.35&2143.66&2141.01&2146.74 \\
196&2144.84&2146.65&2151.09&2142.76&2148.55&2151.00&2151.76&2148.87&2155.49 \\
198&2151.98&2152.87&2158.68&2149.67&2152.87&2158.97&2159.33&2156.53&2163.79 \\
200&2158.78&2165.98&2165.99&2156.29&2161.52&2166.40&2167.00&2163.21&2171.58 \\
\hline \hline
\end{tabular}
 \end{center}
 \end{table}

\begin{table}[h]
\begin{center}
\caption{\footnotesize Binding energies  for $Z=104$  isotopic
chains with calculation of NL3, NL-SH and NL-Z2 sets. $B_{1},
B_{2}$ and $B_{3}$ denote the binding energies of oblate, moderate
prolate and super-deformed prolate configurations. $N$ is the
neutron number. } \label{} \vspace{0.1cm} \scriptsize
\begin{tabular}{|c|c|c|c|c|c|c|c |c|c|c|}\hline\hline
 &\multicolumn{3}{c|}{NL3}&\multicolumn{3}{c|}{NL-SH}&\multicolumn{3}{c|}{NL-Z2}\\
\hline
 {$N$ } &  $B_{1}$(MeV)& $B_{2}$ (MeV)& $B_{3}$(MeV) & $B_{1}$(MeV) & $B_{2}$(MeV) &$B_{3}$(MeV)&$B_{1}$(MeV)& $B_{2}$(MeV)&$B_{3}$(MeV)\\
\hline
152&1876.43 &1889.25&       &1878.83&1892.57&       &1875.29&1887.03& \\
154&1890.14 &1902.90&       &1892.08&1906.31&       &1889.64&1900.68& \\
156&1903.29 &1915.89&       &1904.97&1919.03&       &1902.92&1913.95& \\
158&1915.93 &1928.35&       &1917.46&1931.35&       &1915.50&1926.77& \\
160&1928.13 &1940.28&       &1929.56&1943.22&       &1927.78&1938.71& \\
162&1939.99 &1951.40&       &1941.38&1953.99&       &1940.00&1950.02& \\
164&1951.44 &1961.15&       &1952.84&1964.21&       &1952.00&1960.09& \\
166&1962.32 &1970.77&       &1963.94&1974.32&       &1963.79&1969.74&  \\
168&1972.60 &1980.32&1976.98&1974.68&1984.01&1978.43&1975.48&1979.02&1978.00 \\
170&        &1989.86&1986.94&1985.01&1993.13&1988.48&1986.82&1988.72&1988.07 \\
172&        &1999.24&1997.96&       &2003.14&1997.96&1997.78&1998.55&1997.77 \\
174&        &2008.53&2005.61&       &2012.32&2006.88&2008.14&2008.02&2007.25 \\
176&        &2017.68&2014.23&       &2021.20&2015.58&2018.38&2017.68&2016.55 \\
178&        &2026.58&2022.61&       &2029.66&2023.68&2024.41&2027.31&2025.90 \\
180&        &2034.82&2030.82&       &2037.56&2031.36&2033.09&2035.74&2034.88 \\
182&        &2042.81&2038.60&       &2045.37&2038.75&2041.30&2044.36&2040.45 \\
184&        &2050.54&2045.89&       &2052.65&2045.97&2049.34&2052.73&2048.21 \\
186&        &2056.40&2052.80&       &2058.30&2051.94&2056.83&2059.16&2056.42 \\
188&        &2061.92&2059.48&       &2063.67&2058.63&2063.97&2065.98&2063.95 \\
190&        &2069.40&2066.02&       &2069.59&2065.06&2071.20&2073.86&2071.44 \\
192&        &2075.42&2072.37&       &2075.05&2071.31&2078.14&2080.90&2078.79 \\
194&        &2081.42&2079.81&       &2079.84&2077.46&2084.85&2087.56&2085.65 \\
196&        &2087.61&2085.75&       &2086.46&2083.17&2091.27&2093.73&2092.12 \\
198&        &2093.75&2091.22&       &2092.22&2098.35&2097.51&2100.33&2098.32 \\
\hline \hline
\end{tabular}
 \end{center}
 \end{table}

\subsection{\emph{Binding energy}}
  In the above subsections, we have discussed the deformation and
root mean square radii of $Z=112, 104$ isotopic chains, and find
that there are three deformed chains for each isotopic chain.
Since there are multi-deformed configurations for the superheavy
nuclei, is there shape coexistence in them? In this subsection, we
will discuss it in detail. The binding energies of the three
deformed chains with the force parameters NL3, NL-SH and NL-Z2 are
listed in table 1 and table 2  for $Z=112,104$ respectively.
$B_{1}, B_{2}$ and $B_{3}$ denote the binding energies of oblate,
moderate prolate and super-deformed prolate configurations
respectively. The binding energies with NL3, NL-SH and NL-Z2 in
the same deformed chain are very close, the difference between
them is not larger than 0.05 percent (about 10 MeV). One of the
important conditions for the shape coexistence is that the
deference of binding energies between two deformed configurations
is very small, usually less than 1 MeV. If it is larger than 1
MeV, the translation between two different configurations is very
difficult and the probability of shape coexistence is very little.
To find the sign of shape coexistence, we compare the binding
energies of different deformed chain with each other, and list the
results in figure 3. The circle denotes the values of
$B_{1}-B_{2}$, triangle is for $B_{2}-B_{3}$ and star is for
$B_{1}-B_{3}$. In figure 3, it is seen that all the values of
$B_{1}-B_{2}$, $B_{2}-B_{3}$ and $B_{1}-B_{3}$ are in the region
$-14.5\sim 8$ MeV. It is obvious that the abstract values of them
decrease with the increasing of the neutron number until to a
certain region ($N\sim 184$) and the trend changes when $N > 184$
for $Z=112$ isotopic chain. For $Z=104$ isotopic chain, the
phenomenon is less clear than that of $Z=112$.  It is interesting
that these values are very close to the fission barriers as high
as $8\sim12$ MeV around the double-shell closure $Z =114,
N=184$\cite{t0,t1,t2}. This means the deformation is important for
the stability of superheavy nuclei. From figure 3, we can also
clearly find in which region the shape coexistence exists and what
kind of shape coexistence is in the long isotopic chain.

  For $Z=112$ isotopic chain, from table 1 and figure 3, we can see
$B_{2}$ is always the largest one  and $B_{1}$ is the smallest
among $B_{1}, B_{2}$ and $B_{3}$ in the region $N\leq 186$ for
NL3, $N\leq 188$ for NL-SH and $N\leq 178$ for NL-Z2 calculation
respectively. It indicates that the moderate prolate deformed
configuration is the ground state of nuclei $N\leq 186$ for NL3,
$N\leq 188$ for NL-SH and $N\leq 178$ for NL-Z2 calculation and
the oblate or superdeformed configuration is the exciting state of
them. The most distinct character is that $B_{3}$ is the largest
one when $N\geq 188$ for NL3, $N\geq 190$ for NL-SH and $N\geq
180$ for NL-Z2 calculation, namely, there is a chain of
super-deformed configurations  which become  the ground state and
are more stable than the other deformed configurations. Although
there are some differences with the different force parameters in
predicting the region of super-deformation, all the parameters'
calculation show there is a more stable super-deformed region in
the $Z=112$ isotopic chain. It agrees with the Bohr and Mottelson'
suggestion\cite{t} that deformation can increase the stability of
the heavy nuclei. But it should be noted that it is not for all
heavy nuclei that deformation can increase the stability. This
phenomenon appears only in special region of superdeformed chain.
Figure 3 also shows that the results with NL3, NL-SH and NL-Z2 set
are different in some details, but the total trend is consistent.
In figure 3, the star is below and far from the dotted line
$\delta E$=-1 MeV, it means that it is impossible for oblate and
super-deformed prolate deformation to coexist in $Z=112$ isotopes
. There is no shape coexistence in the region $N\leq 172$,  for no
symbols in the region 1 MeV $\leq\delta E\leq-1$ MeV as shown in
figure 3. The NL3 results show there maybe exist moderate- and
super-deformed shape coexistence in the region $176\leq N\leq 190$
of $Z=112$ isotopes, for the  value of $B_{2}-B_{3}$ (triangles)
is around 1 MeV or within 1 MeV. Although  the NL-SH  results for
$B_{2}-B_{3}$ (triangles) are larger than  1 MeV in the region
$174\leq N\leq 184$, they are very close to 1 MeV, together with
the values within 1 MeV,  we believe that moderate- and
super-deformed shape coexistence maybe occur in $174\leq N\leq
192$. The NL-Z2 results predict the moderate- and super-deformed
shape coexistence should occur in $174\leq N\leq 186$. It is
interesting that all the regions of shape coexistence predicted by
the parameters NL3, NL-SH and NL-Z2 contain the magic number
N=184. It maybe have some trends that shape coexistence occur
usually at around the magic number. NL3 calculation also shows
moderate prolate and oblate deformation coexist at $N=198, Z=112$,
and NL-Z2 shows moderate prolate and oblate deformation coexist at
$N=188,190$. The calculation of Ref.\cite{d3} predicts
$N=172,Z=112$ exists shape coexistence, but our result does not
indicate there is any shape coexistence.

On the other hand, for $Z=104$ isotopic chain, from table 2 and
figure 3, we can find there is no super-deformed configuration
until $ N\geq 168$ and no oblate deformed configuration when $
N\geq 168$ for NL3, $ N\geq 170$ for NL-SH calculation. $B_{2}$ is
always larger than  $B_{3}$ except $N=198$ for NL-SH calculation.
It may be an anomalism or un-physical solution. Thus, for $Z=104$
isotopic chain, the moderate prolate deformed configuration may be
the ground state, and the  deformation can not increase the
stability in this case at all. In figure 3, the results of NL3 and
NL-SH are very close, the triangles, circles or stars are not in
the region -1 MeV$\leq \delta E \leq$ 1 MeV. It indicates there is
no shape coexistence for $Z=104$ isotopes with NL3 and NL-SH
calculations. But the results that NL-Z2 parameter shows are very
different from those of NL3 and NL-SH sets. It predicts moderate
prolate and oblate deformation maybe coexist at $N=172,174$ and
176, moderate prolate and super-deformation maybe coexist in
$168\leq N \leq 180$ and oblate and super-deformation maybe
coexist at $N=172, 174 $ and in the region $182\leq N \leq 198$.
The shape coexistence also occurs around the magic number $N=174$
and 184.

\section{Summary}
We use the considered axial deformed relativistic mean field
theory to perform systematical calculations for $Z=112$ and 104
isotopic chains with force parameters NL3, NL-SH and NL-Z2 sets.
First, three deformed chains (oblate, moderate prolate and
super-deformed chain) are found for $Z=112$ and 104 isotopic
chains. Second, a super-deformed chain can be the ground states
for $Z=112$ isotopic chain when $N\geq 188$ with NL3, $N\geq 190$
with NL-SH and $N\geq 180$ with NL-Z2 calculation. This confirms
that the deformation can increase the stability of some superheavy
nuclei.  Third, although shape coexistence is a general phenomenon
for superheavy nuclei, it only appears in some special regions of
an isotopic chain, and only some special kinds of shape
coexistence can exist. For $Z=112$ isotopic chain, it is predicted
that moderate prolate and super-deformation coexist in the region
$174\leq N\leq 184$ for NL3 set, $174\leq N\leq 192$ for NL-SH
set, $174\leq N\leq 186$ for NL-Z2 set. There is no other kind of
shape coexistence for $Z=112$ isotopic chain except moderate
prolate and oblate deformation at $N=198 $ for NL3 set, moderate
prolate and oblate deformation  at $N=188,190$ for NL-Z2 set. For
$Z=104$ isotopic chain, there is no shape coexistence with NL3 and
NL-SH calculations. However, the NL-Z2 calculation predicts that
moderate prolate and oblate deformation maybe coexist at
$N=172,174$ and 176, moderate prolate and super-deformation maybe
coexist in $168\leq N \leq 180$ and oblate and super-deformation
maybe coexist at $N=172, 174 $ and in the region $182\leq N \leq
198$. We predict that shape coexistence maybe occur around the
magic number. Finally, it is found that there is shell closure at
$N=184$ ($\beta_{2}\approx 0$) for moderate prolate deformed
chains of $Z=112$ and 104,  at $N=176$ for oblate deformed chain
of $Z=112$,  the position of shell closure have strong parameter
dependence for super-deformed chains of $Z=112$ and 104. It is
confirmed the prediction that there is shell anomalism for oblate
or superdeformed nuclei.

\begin{figure}[ht]
\center \epsfxsize=12cm \epsfbox{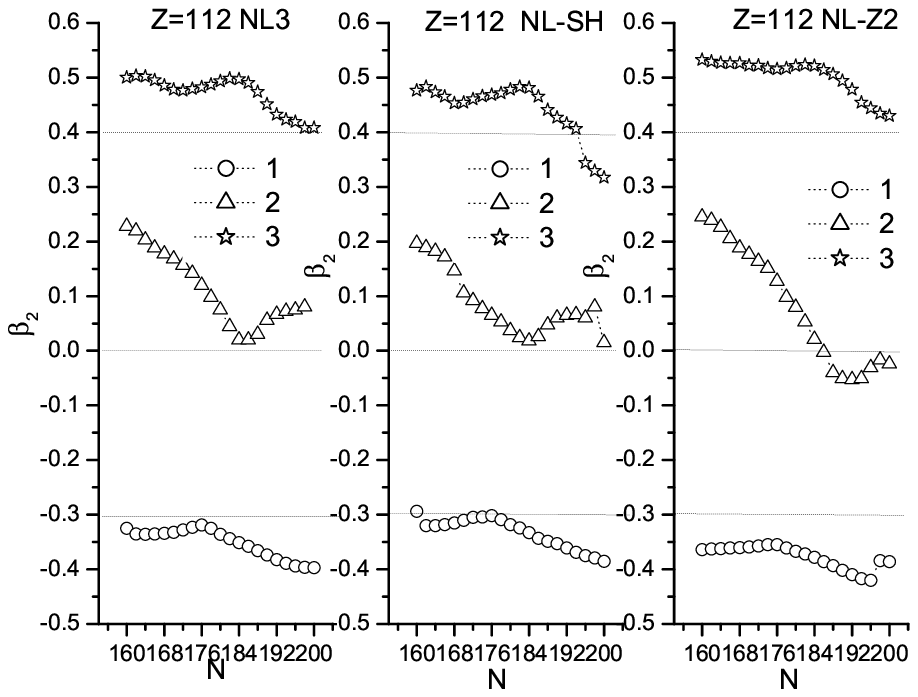} \epsfxsize=12cm
\epsfbox{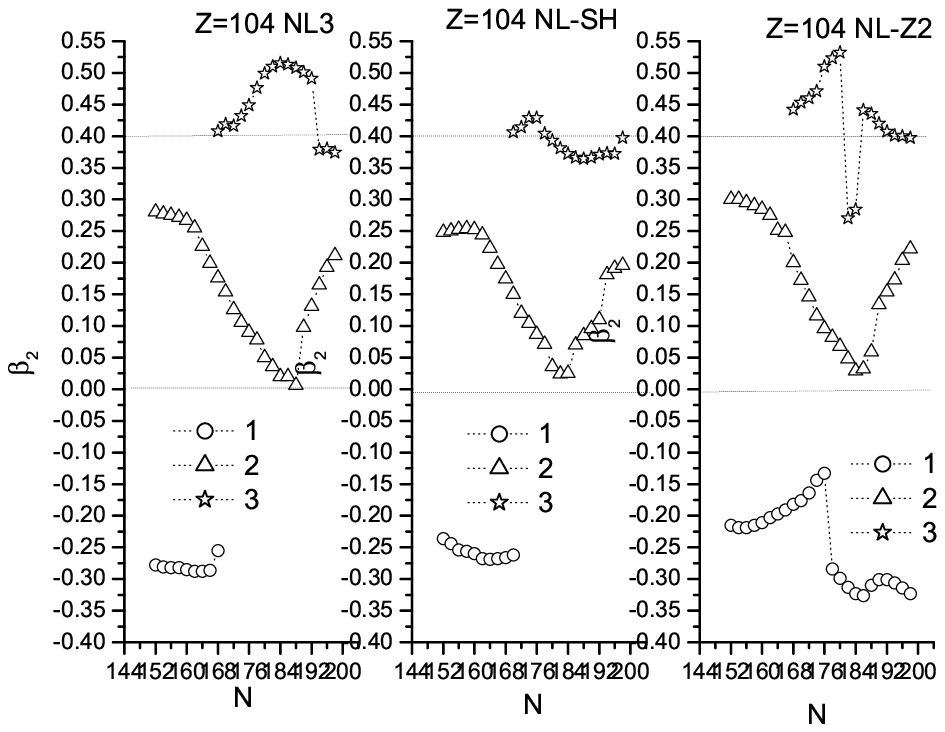} \caption{{ \footnotesize The parameter of
quadrupole deformation $\beta_{2}$ for $Z=112$ and 104 isotopic
chains with calculation of NL3, NL-SH and NL-Z2 sets. Empty
circle, empty triangle and empty star denote oblate, moderate
prolate and super deformation respectively.} }
\end{figure}

\begin{figure}[ht]
\center \epsfxsize=12cm \epsfbox{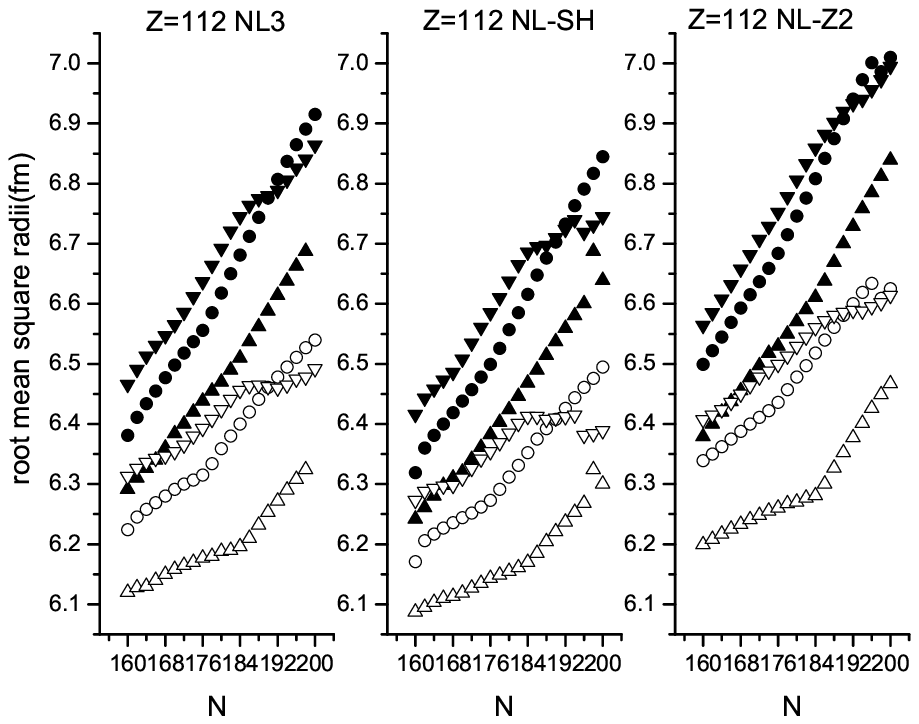} \epsfxsize=12cm
\epsfbox{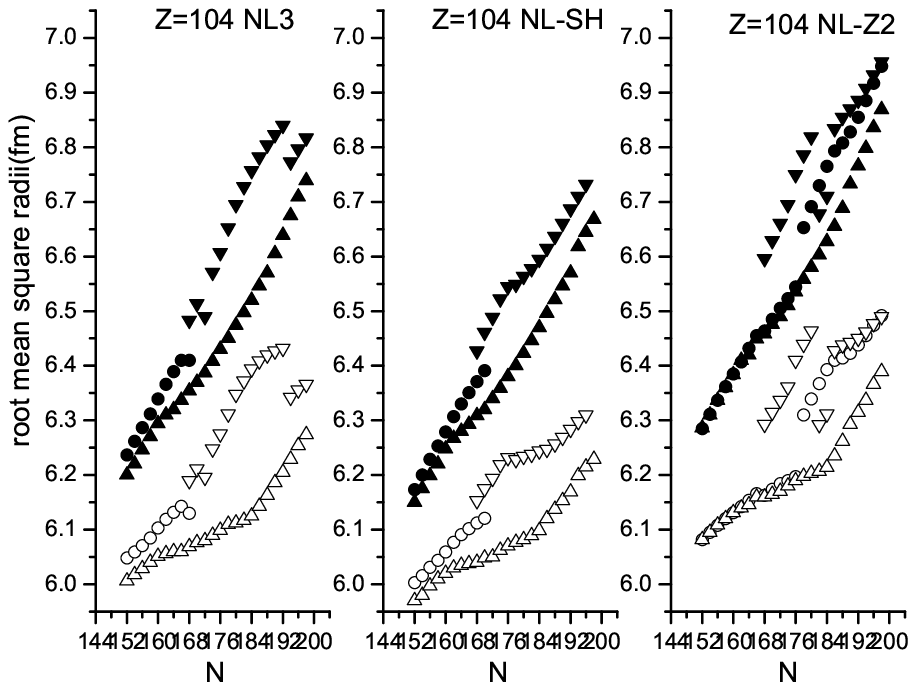} \caption{{ \footnotesize The root mean square
radii for $Z=112$ and 104 isotopic chains with calculation of NL3,
NL-SH and NL-Z2 sets.  Circle, up-triangle and  down-triangle
denote oblate, moderate prolate and super deformed configuration
respectively. The solid symbol stands for neutron and empty symbol
stands for proton } }
\end{figure}

\begin{figure}[ht]
\center \epsfxsize=12cm \epsfbox{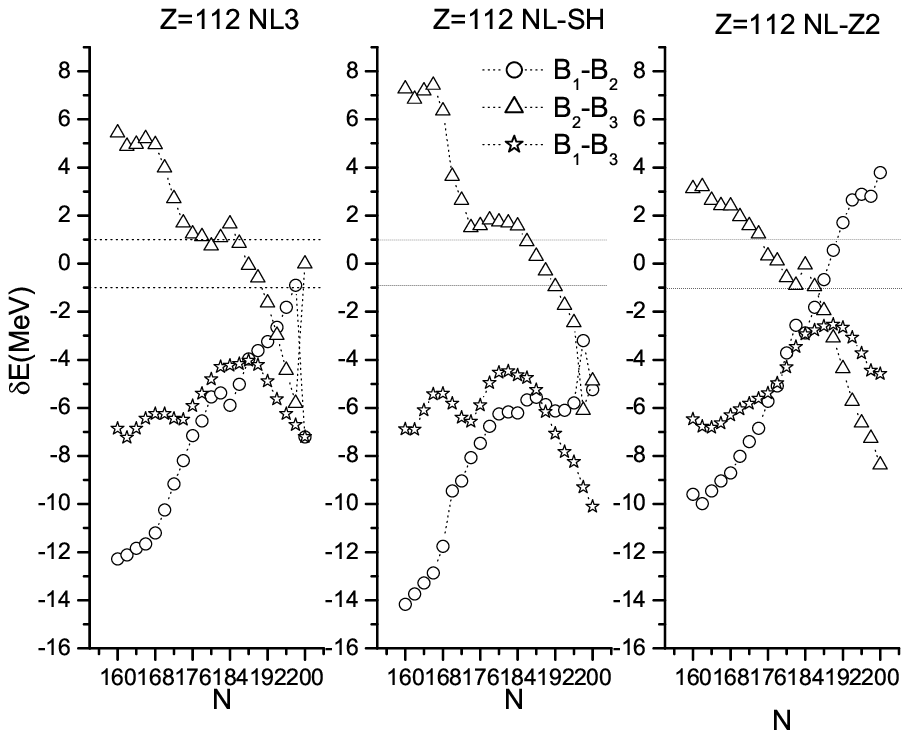} \epsfxsize=12cm
\epsfbox{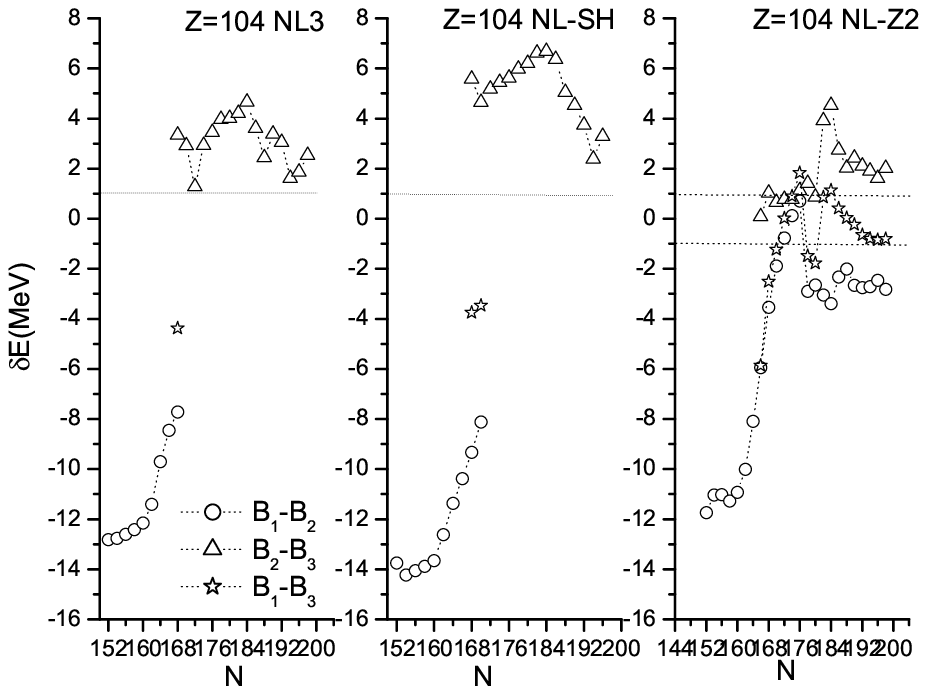} \caption{{ \footnotesize The energy
differences for $Z=112$ and 104 isotopic chains with calculation
of NL3, NL-SH and NL-Z2 sets. $B_{1}, B_{2}$ and $B_{3}$ denote
the binding energies of oblate, moderate prolate and
super-deformed prolate configurations. The circle denotes the
values of $B_{1}-B_{2}$, triangle stands for $B_{2}-B_{3}$ and
star stands for $B_{1}-B_{3}$. } }
\end{figure}

\end{document}